\documentclass{jfm}

\usepackage{graphicx}
\usepackage{natbib}

\ifCUPmtlplainloaded \else
  \checkfont{eurm10}
  \iffontfound
    \IfFileExists{upmath.sty}
      {\typeout{^^JFound AMS Euler Roman fonts on the system,
                   using the 'upmath' package.^^J}%
       \usepackage{upmath}}
      {\typeout{^^JFound AMS Euler Roman fonts on the system, but you
                   dont seem to have the}%
       \typeout{'upmath' package installed. JFM.cls can take advantage
                 of these fonts,^^Jif you use 'upmath' package.^^J}%
      }
  \else
  \fi
\fi

\ifCUPmtlplainloaded \else
  \checkfont{msam10}
  \iffontfound
    \IfFileExists{amssymb.sty}
      {\typeout{^^JFound AMS Symbol fonts on the system, using the
                'amssymb' package.^^J}%
       \usepackage{amssymb}%
         \let\leq=\leqslant
         \let\geq=\geqslant 
      }{}
  \fi
\fi

\ifCUPmtlplainloaded \else
  \IfFileExists{amsbsy.sty}
    {\typeout{^^JFound the 'amsbsy' package on the system, using it.^^J}%
     \usepackage{amsbsy}}
    {\providecommand\boldsymbol[1]{\mbox{\boldmath $##1$}}}
\fi

\newcommand\Rey{\mbox{\textit{Re}}}  % Reynolds number
\newsavebox{\astrutbox}
\sbox{\astrutbox}{\rule[-5pt]{0pt}{20pt}}

\newcommand{\abs}[1]{\left\vert#1\right\vert}
\newcommand{\A}{\mathcal{A}}
\title{Local temperature perturbations in the boundary layer in regime of free
viscous-inviscid interaction}

\author[]%
{M.\ns V.\ns K\ls O\ls R\ls O\ls T\ls E\ls E\ls V$^1$%
  \thanks{Email address for correspondence: m.koroteev@gmail.com},\ns
\and I.\ns I.\ns L\ls I\ls P\ls A\ls T\ls O\ls V$^2$}

\affiliation{$^1$Physics and Biology Unit, Okinawa Institute of Science and Technology,
Okinawa, Onna, Tancha 7542, Japan\\[\affilskip]
$^2$Central Aerohydrodynamics Institute, Zhukovsky, Russia, 140180}

\begin{document}

\maketitle

\begin{abstract}
We analyze the disturbed flow in the supersonic laminar boundary layer when local heated
elements are placed on the surface. It is exhibited that these flows are described in
terms of free interaction theory for specific sizes of thermal sources. We construct the numerical solution for flat
supersonic problem in the viscous asymptotic layer in which the flow is decribed by nonlinear equations for
vorticity, temperature with the interaction condition which provides influence of perturbations to the pressure
in the main order.
\end{abstract}

\section{Introduction}
%\noindent 

The present work continues the studies started in 
\citep{Lipatov2006, Koroteev2009},
which were devoted to construction of asymptotic solutions of Navier-Stokes
equations in the regions, in which local heating elements are situated on the
surface of the body. In \cite{Koroteev2009} we demonstrated how this
problem can be solved analytically for small temperature perturbations.

The idea lying behind this approach consists in the utilizing of small temperature perturbations for
the control of separation of the boundary layer and delay of laminar-turbulent transition.
The former question is related to the zero shear stress point in the flow which can alter its
location if one affects the boundary layer by some, not necessarily thermal, perturbations. The latter is related to
the possibility to slow down the flow by means of the same perturbations to decrease local Reinolds number $\Rey$.

The main mechanism which enables to carry out the control of the boundaty layer is alteration
of the pressure induced by that of displacement thickness in the flow from perturbations. The source of
the perturbations, as we said, can have various nature, e.g., variation of curvature of the surface which
can be realized as small humps or irregularities on it. The problems about perturbations in the boundary layer
which are produced by small humps located on the surface in the bottom of the boundary layer
served the object of thorough studies during past several decades \citep{Bogolepov71, Smith, Bogolepov76}

In \citep{Lipatov2006} the general description of problems emerging when local
heating elements are located on the surface was given. In the same work an important similarity between 
the problems in question and those
of flows over the surface with small humps was demonstrated. It was, in part,
shown that the local heating forms an effective hump and the outline of the flow
becomes similar to that of physical humps on the surface \citep{Smith, Bogolepov76}.

The purpose of the present paper is to extend the analysis to more general and more complicated
case of nonlinear perturbations. Linear problems, corresponding to small perturbations $\Delta T << 1$, 
though being novel, only presented the first step in studing
flows in the boundary layer with thermal perturbations. In addition, the nonlinear problem studied here, 
unlike the linear formulations \citep{Koroteev2009, KoroteevPMM},
requires an extensive use of numerical methods and usually nontrivial algorithms for computation.

\section{Formulation of problem for the viscous sublayer}

We consider the uniform supersonic flow over a flat semi-infinite plate
for large Reinolds numbers $\Rey$ when laminar-turbulent transition is absent.
Denote $\Rey=\epsilon^{-2}$ and consider it as a small parameter. The flow is
assumed to be flat and stationary. On the surface of the body a heating element of size $a$ is located
which produces thermal perturbations in the flow.
The size of the element is an important parameter of the problem and discussed below. We assume
that thermal perturbations are not small $\Delta T\sim T\sim O(1)$.

We are interested in constructing asymptotic solutions of
Navier-Stokes equations supplied in this case by the equation for temperature,
in regions where local perturbations of the temperature exist. This construction
is accomplished in terms of the interaction theory by dividing the region in the
vicinity of the leating element by three smaller regions or decks \citep{Neiland69,
Stewartson69, Messiter70}. Sizes of these regions are proportional to
powers of the parameter $Re=\epsilon^{-2}$. The role of these regions consists in
different character of perturbations in them and different influence on the flow as a whole.
For the problem under consideration
the analysis of equations corresponding to each region was fulfilled in \cite{Koroteev2009,
Lipatov2006} and therefore is not described here. Note that this
analysis shows that the flow is described by the regime of free
interaction \cite{Neiland69, Lipatov2006}. In this regime thermal disturbances
induce pressure gradient which influences the flow functions in the main order. That means
that  diffusive, convective, and pressure gradient terms in the Navier-Stokes equations have the same
order. As exhibited in \citep{Lipatov2006} this condition furnishes the following asymptotic
relations $\Delta p\sim O(\epsilon^{1/2})$, $a\sim O(\epsilon^{3/4})$, $\Delta y\sim O(\epsilon^{5/4})$ which
correspond to the known scales of free interaction \citep{Neiland69, Stewartson69}.
Thus, the free interaction regime emerges only for specific sizes of the thermal region.

\begin{figure}
 \center{\includegraphics[height=180pt, width=250pt]{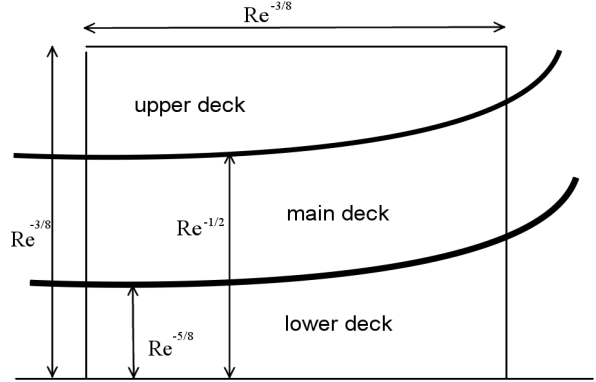}}
 \vspace{0.2cm}
 \caption{Sketch of triple deck region in the vicinity of heating element.}
 \label{fig:TripleDeck}
\end{figure}

From this analysis it also follows that the upper deck is described by inviscid
Euler equations, the middle deck by Prandtl boundary layer equations and the lower 
layer, the closest one to the surface, by some nonlinear set of equations of
parabolic type \citep{Koroteev2009, KoroteevPMM}. The asymptotic layout of three decks is portreyed
in fig. \ref{fig:TripleDeck}. 
In addition,
the equations on the lower deck are supplied by boundary and initial conditions,
which are formulated below, and also by an interaction condition which not only
connects functions of lower deck to those of upper deck, but also provides
influence of the right boundary of the whole flow region to the behavior
of functions inside the region. It implies that the set of equations in the lower deck can
not be viewed as that of parabolic type and gains, in a sense, the property of
ellipticity.

The lower deck is the most essential part of the hierarchy for the pressure
gradient presented in equations for all decks is not fixed but becomes
self-induced, i.e., varies on account of displacement thickness of the boundary layer to
the outer inviscid flow, and the main contribution to the variation of the
pressure is furnished by the very lower deck, while the main deck(the main part
of the boundary layer) remains passive and merely conveys perturbations from the
lower deck to the outer flow. The indicated scheme enables to construct
solutions of equations in the lower deck with self-induced pressure gradient and
thus provides interaction between lower and upper decks.

We give the formulation of the set of equations for lower deck, further called
the viscous sublayer, referring to \citep{Koroteev2009,
KoroteevPMM} for details of derivation.

The set of equations for the viscous sublayer has the form
$$
\frac{\partial u_{b}}{\partial x_{b}}+\frac{\partial v_{b}}{\partial y_{b}}=0,
$$ 

\begin{equation}
\label{2}
u_{b}\frac{\partial u_{b}}{\partial
x_{b}}+v_{b}\frac{\partial u_{b}}{\partial y_{b}}+T_{b}\frac{\partial p_{b}}{\partial x_{b}}=\frac{\partial^{2} u_{b}}{\partial y_{b}^{2}}, 
\end{equation}

$$
u_{b}\frac{\partial T_{b}}{\partial
x_{b}}+v_{b}\frac{\partial T_{b}}{\partial y_{b}}=\frac{\partial^{2}
T_{b}}{\partial y_{b}^{2}}.
$$
Here $u_{b}, v_{b}$ are longitudal and vertical components of velocity in the
sublayer, $T_{b}$ - the perturbation of temperature, $p_{b}(x)$ is the pressure
which only depends on $x$ and consequently is constant through the decks.

The system is supplied by the following boundary conditions. On the surface the
non-slip conditions are stated 
\begin{equation}
\label{slip}
u_{b} (x_{b},0)  =  v_{b} (x_{b},0) = 0.
\end{equation}
We prescribe a temperature profile on he surface:
\begin{equation}
\label{bound_cond_temp1}
T_{b} (x_{b},0)  =  T_{w}(x_{b}).
\end{equation}
The boundary condition at significant distance from the surface is presented as
follows 
$$
u_{b}\to y_{b}+\A(x_{b}), \quad \A_{1}(x_{b})=\int_{0}^{+\infty}(1-T_{b})d\eta +
\A(x_{b}), 
$$
\begin{equation}
\label{bound_cond_temp2}
\quad T_{b}(x,\infty)\to 1,y_{b}\to\infty. 
\end{equation}
The function $\A(x_{b})$ is a standard displacement by means of which the
problems of interaction without heat perturbations are
formulated\citep{JobeBurgraff74, SychevBook}. Local perturbations of the temperature
generate an effective local hump\citep{Lipatov2006} and additionally contribute to
the boundary condition for $y_{b}\to+\infty$, which is expressed by means of a
function $\A_{1}(x_{b})$. The temprature perturbations vanish far from the
location of the heated region and $\A(x_{b})$ tends to $\A_{1}(x_{b})$ as it
follows from the above condition.

The conditions which relate the displacement thickness with the longitudal velocity are
supplied by the interaction condition, which, in turn, relates the displacement thickness
with the pressure which provides the interaction of the viscous sublayer with
the outer inviscid flow.

The condition of interaction differs for subsonic and supersonic flows.
We do not give its derivation for this question was many times discussed in
literature \citep{Stewartson81, SychevBook}. In part, for the supersonic
flow which we study, the condition of interaction has the form

\begin{equation}
\label{interaction}
p(x_{b})=-B\frac{\partial \A_{1}(x_{b})}{\partial x_{b}}.
\end{equation}
Here $B$ is a constant which allows to consider different regimes of
interaction. In our case $B\sim 1$ which corresponds to the regime of free
interaction.

\section{Numerical solution of boundary problem}

The numerical solution the problem (\ref{2}-\ref{interaction}) is constructed on the premise of
the method suggested in \citep{Ruban76}, the detailed
description of which can be found in \citep{SychevBook}. The method was originally employed 
for computations of subsonic problems but can be also utilized for supersonic flows. Supplying the equation
for temperature we thus give but a short
sketch of the procedure, paying attention to necessary modifications of the numerical method.

In the first place the boundary problem described in the previous section
is reduced to that for vorticity $\tau=\partial u/\partial y_{b}$. This is
carried out by differentiating the momentum equation wrt. $y_{b}$ and taking
into account the equation of continuity. Then the system becomes

$$
u_{b}\frac{\partial\tau}{\partial
x_{b}}+v_{b}\frac{\partial\tau}{\partial
y_{b}}+\frac{\partial T_{b}}{\partial y_{b}}\frac{\partial p_{b}}{\partial
x_{b}}=\frac{\partial^{2}\tau}{\partial y_{b}^{2}},
$$
\begin{equation}
\label{vorticity}
\end{equation}
$$
u_{b}\frac{\partial T_{b}}{\partial
x_{b}}+v_{b}\frac{\partial T_{b}}{\partial y_{b}}=\frac{\partial^{2}
T_{b}}{\partial y_{b}^{2}}.
$$
The boundary conditions have the following form
\begin{equation}
\label{bound_cond1}
\tau\to 1, y_{b}\to+\infty, x_{b}\to -\infty.
\end{equation}
Non-slip conditions (\ref{slip}) as well as those for the temperature
(\ref{bound_cond_temp1}) are retained on the wall. 

Next it is easily seen that the momentum equation gives
\begin{equation}
\label{boundCond2}
T_{w}(x_{b})\frac{\partial p}{\partial x_{b}} =
\frac{\partial\tau}{\partial y_{b}}, \quad y_{b}=0.
\end{equation}
From this equation we obtain using the interaction condition (\ref{interaction})
\begin{equation}
\label{interaction2}
-BT_{w}(x_{b})\frac{\partial^{2} \A_{1}}{\partial x^{2}_{b}} =
\frac{\partial\tau}{\partial y_{b}}, \quad y_{b}=0.
\end{equation}

Localization of the effective hump implies the decay of perturbations downstream, namely
$$
p(x_{b})\to 0, \quad x_{b}\to+\infty.
$$
The last condition implies, according to (\ref{interaction}), that
\begin{equation}
\frac{\partial \A_{1}(x_{b})}{\partial x_{b}}\to 0, \quad x_{b}\to +\infty.
\end{equation}

The numerical method provides
some freedom in the consecutive evaluations of the functions as the
number of relations between them exceeds the number of functions. We describe but one possible realization of
the procedure which 
enables to carry out the computations.

The grid is given by $(x_{j},y_{k})$, $j=1,2\ldots N, k=1,2\ldots M$, with
steps $\Delta x, \Delta y$ respectively.

The system (\ref{vorticity}) is approximated as follows
$$
u_{jk}\left(\frac{\partial
\tau}{\partial x}\right)^{*}_{jk} + v_{jk}\frac{\tau_{jk+1}-\tau_{jk-1}}{2\Delta
y} + p^{\prime}_{jk}\frac{T_{jk+1}-T_{jk-1}}{2\Delta y}=
$$
$$
=\frac{\tau_{jk-1}-2\tau_{jk} +\tau_{jk+1}}{\Delta y^{2}},
$$
\begin{equation}
\label{discrete}
\end{equation}
$$
u_{jk}\frac{T_{j+1k}-T_{jk}}{\Delta x} +
v_{jk}\frac{T_{jk+1}-T_{jk-1}}{2\Delta y} + \frac{T_{jk-1}-2T_{jk} + T_{jk+1}}{\Delta y^{2}}
$$
where
\begin{equation}
    \left(\frac{\partial\tau}{\partial x}\right)^{*}_{jk}= \left\{
   \begin{array}{ll}
     \frac{\tau_{jk} - \tau_{j-1k}}{\Delta x}, & u_{jk} \geq 0\\[2pt]
     \frac{\tau_{j+1k} - \tau_{jk}}{\Delta x}, & u_{jk} < 0
   \end{array} \right.
\end{equation}

We thus notice that depending on the sign of the longitudal velocity different
templates are employed. This is quite a famous method for securing static stability of
the difference scheme \citep{Roache}. Derivatives wrt. $x$ can be approximated using the values
in three points \citep{SychevBook}. However our computations show that this alteration does not produce
noticeable effect in accuracy, at least for supersonic flows. Therefore we employed tow-point templates.

The solution is constructed by iterations with respect to $\A_{1}(x_{b})$ using
under relaxation \citep{SychevBook}. Let the approximation of $\A_{1}^{n}(x_{b})$ is
known on some step of the iteration procedure. At each line $x=x_{j}$ the first
equation of (\ref{discrete}) is reduced to a linear set of algebraic equations
with a tridiagonal matrix and solved consecutively from the bottom to the top of
the grid with the boundary conditions (\ref{interaction2}) and
(\ref{bound_cond1}) and then the equation for temperature (\ref{discrete}) is solved
with the boundary conditions  (\ref{bound_cond_temp1}), (\ref{bound_cond_temp2}). Then
the computations are transferred to the next line. It  allows to fulfill computations of
vorticity and temperature in the whole 2D region.

\begin{figure}
 \label{fig:error}
 \center{\includegraphics[height=160pt, width=240pt]{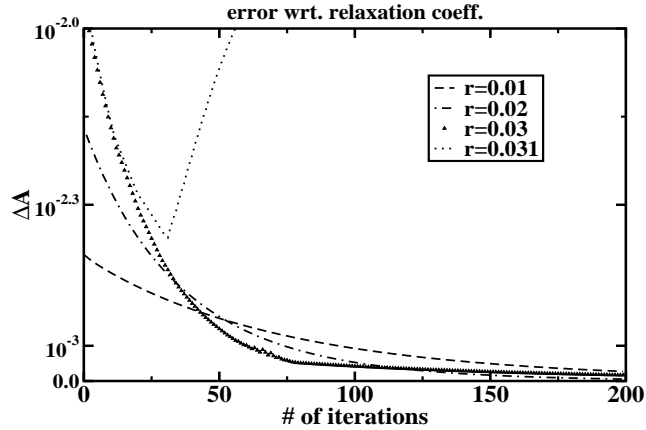}}
 \vspace{0.2cm}
 \caption{Behavior or error $\Delta A$ in consecutive iterations for various values of the relaxation parameter $r$.}
 \label{error}
\end{figure}

The right boundary attained, the functions $u_{jk}, v_{jk}$ as well as $\hat{\A}_{1}(x_{b})$ are recomputed 
on the current iteration in the whole field and
the new approximation for $\A_{1}(x_{b})$ is evaluated from the relaxation
$$
\A_{1}^{n+1} = (1-r)\A_{1}^{n} + r\hat{\A}_{1}(x_{b}),
$$
where $r$ is a relaxation parameter. In problems of interaction theory it usually turns out that
convergence of iterations is strongly sensitive to the values of this parameter\citep {Smith77, SychevBook}.
The problem under consideration is not an exception. In fig. \ref{fig:error} 
the errors $\Delta A=\A^{n+1} - \A^{n}$
are presented for several values of the relaxation parameter. It is noticed that convergence, which exists 
for $r=0.02$, vanishes even for $r=0.031$. In our computations we took $r=0.02$, the value which was found empirically
and which yields an estimate of the upper boundary of some interval of convergence.

\section{Results of computations}

The key functions to evaluate are distributions of pressure and shear stress on the surface.
As we said, the self-induced pressure is constant through the layers and thus its values remain valid in the inviscid
flow for the fixed $x_{b}$. The shear stress in turn can be thought of as a general function characterizing forces in the flow.

\begin{figure}
\center{\includegraphics[height=180pt, width=250pt]{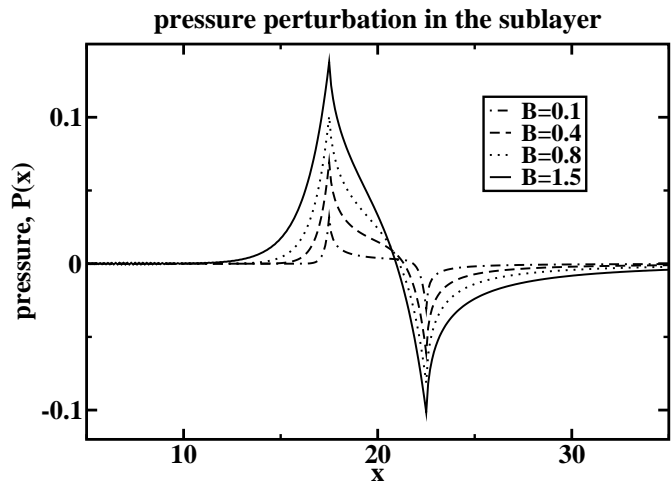}} 
\vspace{0.2cm}
\caption{Pressure in the viscous sublayer obtained from the numerical solution of the problem. Small $B$ decline pressure perturbations even if $\Delta T$ is not small.}
\label{fig:pressure}
\end{figure}

In fig. \ref{fig:pressure} it is exhibited the pressure in the viscous sublayer for heat humps,
of a somewhat artificial form, which are given by 
 
\begin{equation}
\label{hump}
     T_{w}(x_{b})= \left\{
    \begin{array}{ll}
      1 + h, & \abs{x_{b}}\leq a/2\\[2pt]
      1, & \abs{x_{b}} > a/2
    \end{array} \right.
\end{equation}
Here $h$ the perturbation of the temperature and $a$ the size of the heated region. For computations
we take $h=0.3-0.4$.

This form of temperature distribution is simpler for analytic study and in the same time
enables to represent adequately qualitative behavior of functions in the flow. The results are presented
for various $B$. It is noticed that the pressure perturbations decrease for small $B$ even if $\Delta T\sim O(1)$. This 
situation corresponds to compensated interaction regime which is discussed elsewhere. 

\begin{figure}
 \vspace{1cm}
 \center{\includegraphics[height=180pt, width=250pt]{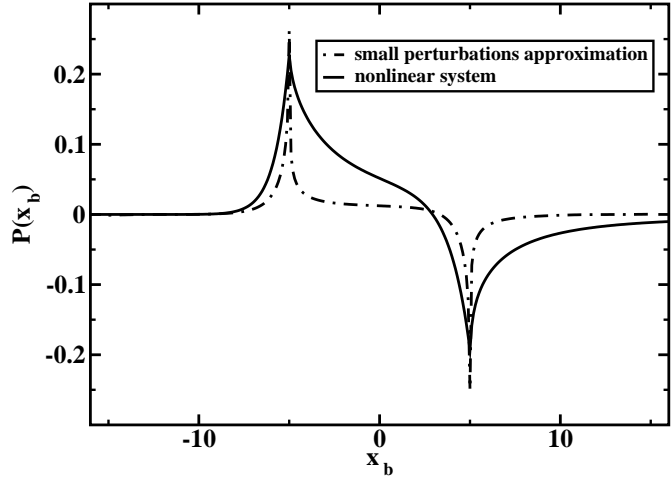}}
 \vspace{0.2cm}
 \caption{Pressure in the viscous sublayer obtained from the numerical solution of the problem compared to linear solution which corresponds to small
 temperature perturbations and was presented in \citep{Koroteev2009}. For both cases the heated region is determined by (\ref{hump}) with the size
 $10$.}
 \label{fig:pressure_comp}
\end{figure}

The comparison of computations for
nonlinear and linear cases, the latter corresponding to small temperature perturbations, is represented in fig. \ref{fig:pressure_comp}. 
The difference is especially noticeable in the transient region and downstream which are not quite adequately approximated by the linear theory.

\begin{figure}
 \center{\includegraphics[height=180pt, width=250pt]{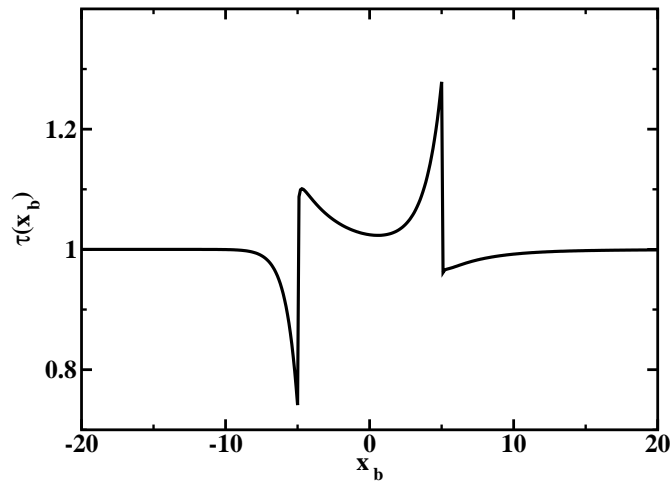}}
 \vspace{0.2cm}
 \caption{Shear stress in the boundary layer.}
 \label{fig:shear}
\end{figure}

The flow moves
from the left to the right. When approaching the heated region the positive pressue gradient produces
aditional deceleration of the flow, thus delaying possible laminar-turbulent transition. Note that there are
two possible separation points in the flow, namely boundaries of the heated region upstream and downstream (fig. \ref{fig:shear}). They have
to be controlled if the separation is not desirable. In the transient region between these points and over
the heating element the flow accelerates affected by the negative pressure gradient which simultaniously increases the shear stress.
Finally perturbations of the pressur decay Behind the heated region and the flow againg slows down.
The parameter $B$ from (\ref{interaction2}) influences the amplitude of pressure perturbations. As we mentioned
before the regime of free interaction implies $B\sim 1$.

\section{Conclusions}

The problem whose solution is presented in this paper presumabily can essentially contribute to possibility of
control of boundary layer and general understading of influence of thermal perturbations to both subsonic and supersonic flows.
Possible closest developements of this work imply generalization of the obtained solutions to three dimensional case which is
of course of more interest for applications. From the other hand, another direction would be the generalization of these results
to nonstationary flow where it is essential to study nonlinear perturbations and their propagation. Both these problems are under
consideration and will be published elsewhere.

\bibliographystyle{jfm}
\bibliography{hydrodynamics}

\end{document}